\documentclass[12pt]{iopart}

\usepackage{iopams} 
\usepackage{epsfig}

\newcommand{\be}{\begin{equation}}
\newcommand{\ee}{\end{equation}}
\newcommand{\lp}{\left(}
\newcommand{\rp}{\right)}
\newcommand{\bra}{\langle} 
\newcommand{\ket}{\rangle}
\newcommand{\half}{\frac{1}{2}}

\begin{document}

\title[Path-phase duality with translational-internal entanglement]{Path-phase duality of an
interfering particle with translational-internal entanglement}

\author{Michal Kol\'{a}\v{r}$^{1,2}$, Tom\'a\v s Opatrn\'y$^1$, Nir Bar-Gill$^2$, Noam Erez$^2$ and Gershon Kurizki$^2$}

\address{$^1$ Department of Optics, Palack\'{y}
University, 17. listopadu 50,
77200 Olomouc, Czech Republic\\ $^2$ Weizmann Institute of Science, 76100
Rehovot, Israel}
\ead{kolar@optics.upol.cz}

\begin{abstract}
The aim of this paper is to revisit the implications of complementarity when we inject into a Mach Zehnder interferometer 
particles with internal structure, prepared in special translational-internal entangled (TIE) states.
This correlation causes the path distinguishability to be phase dependent in contrast to the standard case. 
We show that such a TIE state permits us to detect small phase shifts along with almost perfect path distinguishability, beyond the constraints imposed by  
complementarity on simultaneous which-way and which-phase measurements for standard cases (when distinguishability is independent of interferometric phase).
\end{abstract}


\section{Introduction}

Wave-particle duality is one of the basic features of quantum mechanics: particles sent through an interferometer can produce wave-like interference fringes, but once we try to find which path the particle has taken, the fringes disappear \cite{Bohr,Zurek,Feynman,ScullyNature}. A more detailed analysis has identified the complementarity relating 
our knowledge of the particle's path and the fringe visibility \cite{Wootters}.
The simplest complementarity relation can be expressed in the following form \cite{Wootters, Jaeger, Englert, RempeNature, Luis, Kim, Duerr, Kasevich}, which has been
experimentally verified \cite{RempeNature}
\begin{eqnarray}
 D^2 + V^2 \le 1,
\label{e1}
\end{eqnarray}
where $D$ is the path distinguishability, and $V$ is the fringe visibility. Relation (\ref{e1})
was derived under assumption that $D$ is independent of the interferometric phase. This is what we have
in mind while referring to a ``standard'' case, scheme, etc. 

Consider first ``standard schemes''. In standard schemes, which-path (henceforth which-way--WW) information is obtainable either if the alternative paths have a priori unequal detection 
probabilities \cite{Wootters} or if a WW detector of finite efficiency is placed along  
the paths \cite{Jaeger,Englert}. Clearly, these options may be combined. 

Now suppose the phase difference between the alternative paths is allowed to assume only two possible values, $\phi_0\pm|\delta \phi|$. The \emph{joint}
probability of correct which-way (WW) and which-phase (WP) guesses is constrained by (\ref{e1}) for \emph{any} detector efficiency and phase difference, $\delta \phi$.
The standard constraints on this specific task are discussed in Sec.\ref{comp} for either mono- or poly-chromatic beams in a Mach Zehnder interferometer (MZI). 
The aim of this paper is to show that the standard constraints on this task can be avoided if we inject into the MZI particles with internal structure, prepared in special 
translational-internal entangled (TIE) states introduced in Refs.\cite{game06,QI06} (Sec.~\ref{model}). 
We show that such a TIE state permits us to detect  small phase shifts along with almost perfect path distinguishability, beyond the constraints imposed by standard 
complementarity, on simultaneous WW and WP information (Sec.~\ref{s2.1} and \ref{s2.2}, compared to Sec.~\ref{WW&WP}).
We stress that entanglement in a TIE state refers to intraparticle correlations of inseparable
degrees of freedom, rather than to interparticle entanglement, so the Bell inequalities or
nonlocality play no role. 

Finally, in (Sec.~\ref{comptie}) we show that standard complementarity relation (\ref{e1}) is not violated and introduce a relation between phase sensitivity of the interference pattern and the path distinguishability. For particles without internal 
structure, this relation is in agreement with (\ref{e1}), but for particles prepared in TIE states it allows for much higher simultaneous accuracy of path and phase guesses 
than the accuracy permitted by (\ref{e1}) (Sec.~\ref{sec6}). These results are discussed in Sec.~\ref{sec7}.

\section{Standard complementarity in a Mach Zehnder interferometer (MZI)}

\label{comp}

\numparts

The likelihood $P_{\rm WW}$ of guessing correctly the outcome of a ``which-way'' (WW) measurement (assuming the optimal strategy)
may be related to $D$, the ``distinguishability'' or ``detection efficiency'' of the detector states \cite{Englert}, via:
\be
P_{\rm{WW}}=\frac{1+D}{2} \label{P_ww}.
\ee 

The fringe visibility is recorded by scanning the detection probabilities at the two MZI output detectors  
denoted, e.g., by $+$ and $-$,  across the range of possible phase differences $\phi$:

\begin{figure}[tbh]
\centering \hspace{-0.06\linewidth}
\includegraphics[width=.6\linewidth]{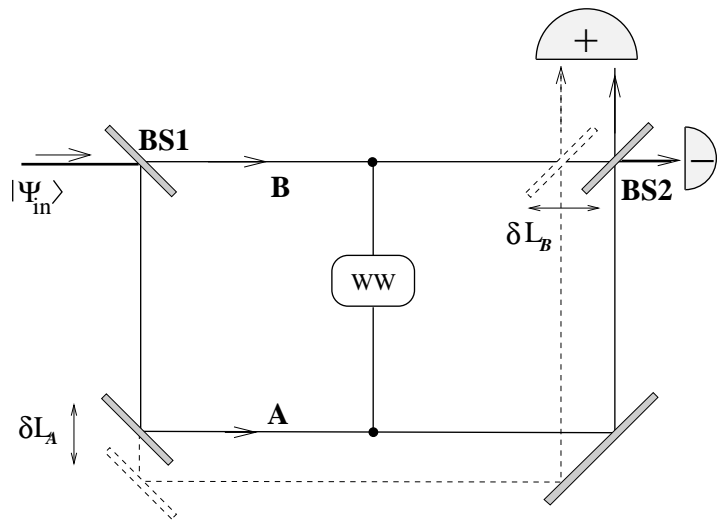}
\caption{Standard MZI scheme (with WW detector coupled to the arms) and output detectors $+$ and $-$.
BS1 and BS2 are balanced (50\%-50\%) beam splitters. Path difference $L_{0}$ may deviate by $\pm|\delta L|=\lp \delta L_A -\delta L_B\rp $
(see Sec. \ref{s2.1}).} \label{f1}
\end{figure}

\be
P_{\pm}=\frac{1}{2}(1\pm V \cos \phi).
\ee

The fringe visibility, $V$, can be similarly related to a ``which phase'' \cite{Kim} (WP)  
probability for correctly guessing the output phase to be $\phi_1$ or $\phi_2$.
For two \rm{orthogonal} output states, labeled by $\phi_1=0,~~\phi_2=\pi$,

\be
\lp P_{\rm{WP}}\rp_{\rm{max}}=\frac{1+V}{2}. \label{wp_prob}
\ee

\endnumparts

The probabilities $P_{\rm{WW}}$ and $\lp P_{\rm{WP}} \rp_{\rm{max}}$ obey the following inequality inferred from (\ref{e1}):

\be
\lp 2P_{\rm{WW}}-1 \rp^2 + \left[ 2\lp P_{\rm WP}\rp_{\rm{max}} -1 \right]^2 \leq 1. \label{Duality2}
\ee

A particle propagating in a standard MZI, arriving at the beam merger, having interacted with a WW detector, 
is in an entangled state with it: 
\numparts
\be
|\Psi_{\rm{st}}\ket = c_1|A\ket|d_A\ket+c_2|B\ket|d_B\ket
\ee
where $|A\ket$ and $|B\ket$ are the path states and $|d_A\ket$ and $|d_B\ket$ are the corresponding detector states (for simplicity, it is assumed that the detector
is initially in a pure state). We will be particularly interested in the case where $|c_1|^2=|c_2|^2=\frac{1}{2}$, obtainable for balanced (50\%-50\%) beam splitters,
so that the a priori visibility is 1, and is reduced only by the presence of the detector. 
Under these assumptions, the complementary measures $D$ and $V$ can be expressed in terms of the 
overlap of the alternative detector states \cite{Englert}:

\begin{eqnarray}
D &=& \lp 1 - |\bra d_A| d_B\ket|\rp^{1/2}  \label{e4a}      \\
V &=&  |\bra d_A|d_B\ket|.  \label{e4b}
\end{eqnarray}

\endnumparts

\subsection{Complementarity for discrete WP and WW outcomes}
\label{s2.1}

Now consider the case where the length of one of the
interferometer arms can assume two discrete values, changing the path-difference by $+|\delta L|$ or $-|\delta L|$, hence the interferometer is unbalanced (Fig. \ref{f1}).
We would like to quantify the tradeoff between the WW probability as defined above, 
and the WP probability of distinguishing between the two possible path- or phase-differences. We now 
assume the input state to be a plane wave $e^{i k x}$ (as an approximation to a realistic input--see Sec.~\ref{model}).
Let $L_{A,B}$ denote the lengths of the two interferometer arms, and let the
two possible path length \rm{differences} be $L_{AB}=L_A-L_B=L_0 \pm |\delta L|$.

If the corresponding phase difference is not an odd multiple of $\pi/2$
($k L_0=n\pi/2$, but $~~k \delta L\neq n'\pi/2$, $n$ and $n'$ 
being odd integers), the results (\ref{wp_prob}) and (\ref{Duality2}) are modified. Namely, 
the probabilities for detection at the two output ports (in Fig.~\ref{f1}) are now, respectively:

\numparts
\be
P_\pm = \frac{1}{2} \pm \frac{V}{2} \cos{\left[k\lp L_0+ \delta L\rp\right]} \label{e5a}.
\ee
Without loss of generality, let us assume that for positive $\delta L$, $P_+>P_-$.
Then the probability of guessing correctly the sign of the phase $\pm \delta \phi = \pm k|\delta L|$ is maximized if we associate detection 
at port $+$ (or $-$) with the path-difference change $|\delta L|$ (or $-|\delta L|$). Hence

\be
P_{\rm WP}\lp \delta L \rp=P_+=\frac{1}{2}+\frac{V}{2}\sin k|\delta L|
\ee which differs from $\lp P_{\rm WP} \rp_{\rm{max}}$ in (\ref{wp_prob}).
\endnumparts
The duality relation, \rm{as inferred from} (\ref{e1}), now takes the form 

\be
\lp 2P_{\rm WW}-1 \rp^2 + \lp \frac{2P_{\rm WP}(\delta L) - 1}{\sin k |\delta L|} \rp^2 \leq 1. \label{Duality3}
\ee 
This \rm{differs} from Ineq.~(\ref{Duality2}). When $k \delta L =0 $, Ineq.~(\ref{Duality3}) implies $P_{\rm WP}=\frac{1}{2}$.

\numparts

We may \rm{always} (formally) parameterize the detector efficiency (\ref{e4a}) as 

\be
D=|\cos \alpha|. \label{alpha}
\ee 
If the detector efficiency is very close to 1, i.e. $\alpha\ll 1$  
\be
P_{\rm WW} \simeq 1-\frac{\alpha^2}{4}, \label{9}
\ee and $k|\delta L| \ll 1$, then Ineq.~(\ref{Duality3}) may be written as:

\endnumparts

\be
P_{\rm WP} \lesssim \frac{1}{2}\left[1+|\alpha|\sin{(k|\delta L|)}\right]\simeq \frac{1}{2}\lp 1+ k |\alpha\delta L |\rp. \label{e9}
\ee
Let us point out that (\ref{e9}) was derived under ``standard'' assumptions, i.e. $D$ is independent of $\phi=kL_{AB}$.

%

\subsection{Generalization to polychromatic wavepackets} \label{s2.2}

We may generalize these considerations to the case that a \emph{polychromatic wavepacket} is used as 
the input state of the MZI (right before impinging on beam splitter (BS1)):
\numparts
\be \label{eqs9}
\bra x| \psi (t) \ket =\sum_{j=1}^N c_j e^{i\lp k_j x -\omega_j t \rp} \equiv f(x,t).
\ee
Inside the MZI, (right before beam merger (BS2)) the \emph{combined state} of the system+detector is then:

\be
|\Psi_2(t) \ket = \frac{1}{\sqrt{2}} \lp f(L_A, t) |A\ket | d_A \ket +  f(L_B, t) |B\ket | d_B \ket\rp.
\ee
For sufficiently long detection times, upon averaging out terms oscillating as $e^{i(\omega_j-\omega_j')t}$, we then find, instead of (\ref{e5a}), the 
time-averaged detection probabilities

\be
\bra P_\pm \ket_T = \half \left\{ 1 \pm \tilde V  \sum_{j=1}^N |c_j|^2\cos \lp k_j L_{AB}+ \phi \rp  \right\} \label{e9c}
\ee where we have defined (cf. Eq.~\ref{e4b}).

\be
\tilde V = |\bra d_A| d_B \ket| = \sqrt{1-D^2}.
\ee
\endnumparts
The actual visibility, $V$, is less than or equal to $\tilde V$. Hence, duality as 
expressed by (\ref{Duality2}) or (\ref{Duality3})
 is perfectly adequate for polychromatic wavepackets.
This will be contrasted below with the results for TIE states.

\section{TIE states in MZI}
\label{model}

\begin{figure}[tbh]
\centering
\includegraphics[width=.8\linewidth]{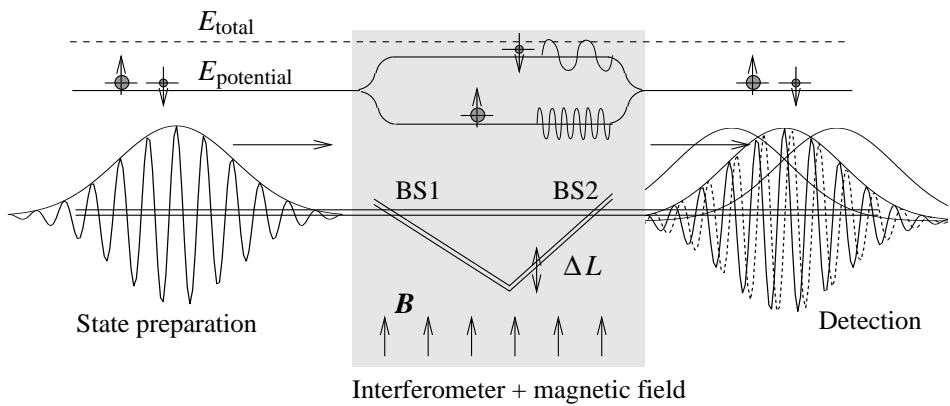}
\caption{
Scheme for state (\ref{e2}) preparation and its passage through an interferometer. The upper part shows the energy dependence on position. An atom is initially prepared in a superposition of spin up and spin down states. Upon entering the magnetic field region, the potential energy of the atom is split depending on the spin state: spin-down atoms are slowed down whereas spin-up atoms are accelerated. The lower part shows the interferometer where the paths are realized by atomic waveguides, e.g., red detuned optical beams. The initial wavepacket is split into two components moving with two different speeds inside the magnetic field region, whereas their speeds before and after this region are equal. The wavepacket width must be sufficiently large for these two components to overlap after leaving the interferometer.} \label{f2a}
\end{figure}

\numparts
Consider spin-1/2 particles (or their analogs: two-level atoms) of
mass $M$ that are prepared in the entangled input
state
\begin{equation}
|\psi _{\mathrm{input}}\rangle =c_1 |k_1
\rangle |1\rangle + c_2 |k_2 \rangle |2\rangle . \label{e2}
\end{equation}
Here the internal states $| 1 \rangle,| 2 \rangle$ correspond to the
internal energy levels $\epsilon_1,\epsilon_2$ and $\hbar k_1,\hbar
k_2$ are $x$-oriented momenta, constrained by the total energy of
state (\ref{e2}):
\begin{equation}
E = \frac{\hbar^2 k_1^2}{2 M} + \epsilon_1 = 
\frac{\hbar^2 k_2^2}{2 M} + \epsilon_2. \label{e1b}
\end{equation}
Condition (\ref{e1b}) ensures that no \emph{time-oscillations} are associated with a superposition of $|k_1\ket|1\ket$ and $|k_2\ket|2\ket$.
Such a state can be created, e.g., using the longitudinal Stern Gerlach effect (see Fig. \ref{f2a}). To this end, cold atoms are prepared in a superposition 
of two spin states, $c_1 |1\rangle +  c_2 |2\rangle$. The states $|1\rangle$ and $|2\rangle$ correspond to two orthogonal spin orientations along the $z$ axis. 
The atoms then enter a region with $z$-aligned magnetic field where they can move along two alternative $x$-oriented paths forming the MZI. The atoms are confined 
to one-dimensional motion along those paths by atomic waveguides: narrow beams of light red-detuned from an atomic transition \cite{Kasevich}. 
Similar elements form the beam splitter and merger. The magnetic field gradient accelerates or decelerates atoms whose spin is parallel or antiparallel to the field, respectively. By adjusting the field intensity  and the initial speed of the atoms we can control the ratio between the momenta $\hbar k_1$ and $\hbar k_2$.  Since the potential is conservative, the solution of the Schroedinger equation is \emph{time independent} and the spin precession about the $z$ axis is locked to the motion along the $x$ axis. Thus, the spin orientation is uniquely determined by the position, i.e., at a distance $x$ from the point of entrance into the magnetic field the internal state of the atom is 
\be
|\psi (x) \rangle = c_1\exp{(\rmi k_1x)}|1\rangle + c_2\exp{(\rmi k_2 x)}|2\rangle. \label{e10c}
\ee 
\endnumparts

When the atoms leave the  magnetic field region, the gradient slows down the fast atoms and accelerates the slow atoms, so that their final speed  is equal to the initial one.
As the spin stops precessing in the field-free zone, a measurement of the spin orientation in this zone  can give us information about the distance the atom has traveled 
in the magnetic field region. 

\Eref{e2} is a plane-wave idealization of realistic wave packets whose spatial width $\Delta x$ is the inverse of the momentum spread, $\Delta x \Delta k \sim 1$. In the magnetic field, the wave-packet is split into fast- and slow- moving components whose overlap is gradually diminishing. If the spin state is to be used for measurement of the distance traveled by the atom, $\Delta x$ must be sufficiently large, so that the two components substantially overlap after leaving the magnetic field.

An alternative realization of atomic TIE states and their propagation in a MZI may be based on 
elements of the scheme in \cite{RempeNature}: atomic Bragg reflection, beam splitting and momentum entanglement with hyperfine levels by standing wave laser beams.

The effect of spin rotation in (\ref{e10c}) is similar to the photon polarization rotation along the different interfering paths in \cite{Japha}, where the 
Faraday effect was used as a ``quantum clock'' for measuring traversal times of photons.

\section{Which-way and which-phase measurements for TIE}
\label{WW&WP}

\begin{figure}[tbh]
\centering \hspace{-0.06\linewidth}
\includegraphics[width=.6\linewidth]{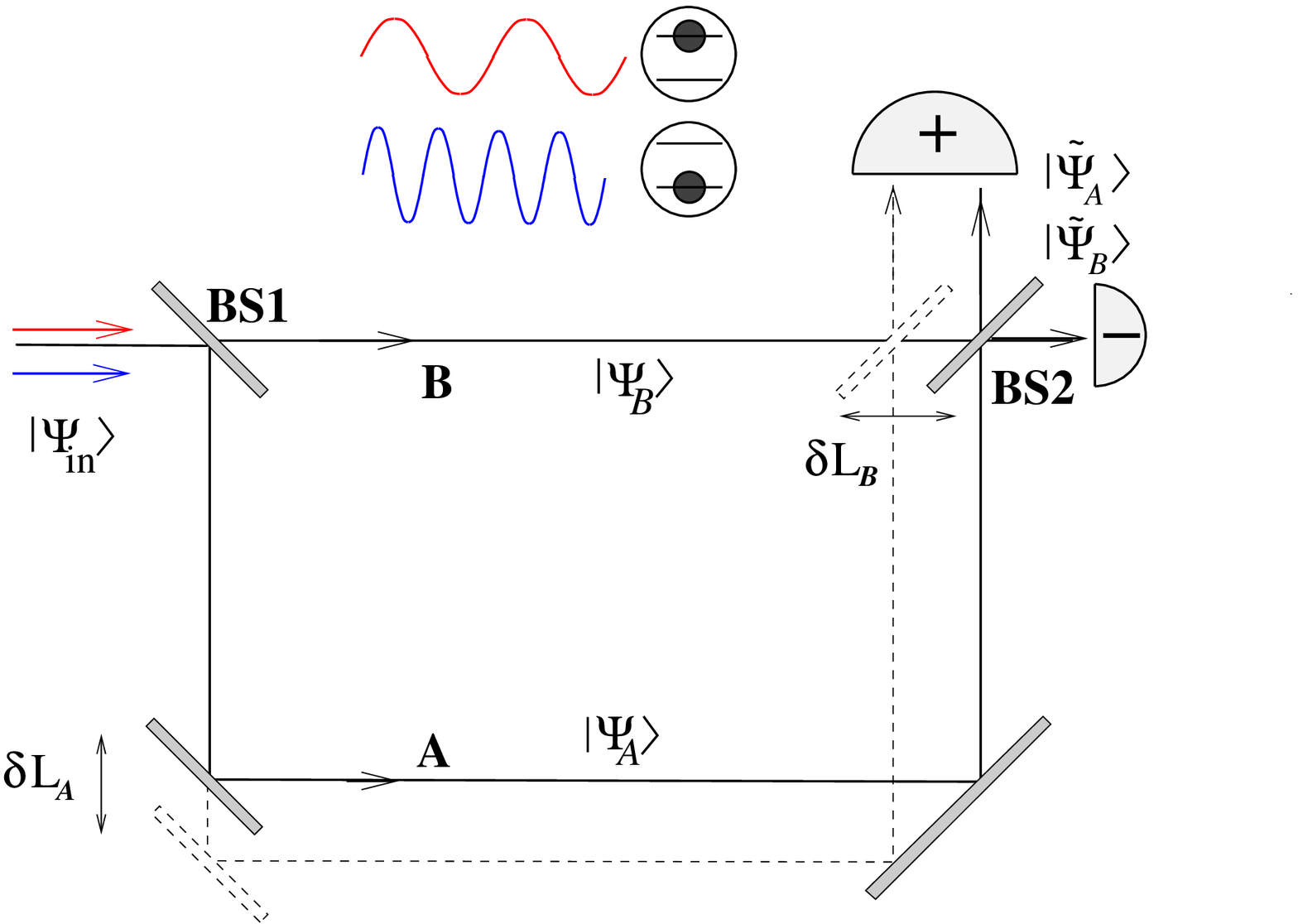}
\caption{ A particle in the state (\ref{e2}) in a MZI. It
traverses the interferometer from BS1 to BS2 via paths A and B,
whose length difference $L_{A0}$, $L_{B0}$ is changed by $\delta L_A$, $\delta L_B$, respectively. Both output detectors
$+$ and $-$ discriminate internal states $|\tilde{\psi}_A \rangle$ and $|
\tilde{\psi}_B \rangle$ (Eq.(\ref{e11a})) (Compare with Fig.\ref{f1}).  
} \label{f2b}
\end{figure}

Particles prepared in state (\ref{e2}) exhibit unusual behavior in a MZI (Fig.~\ref{f2b}).
The incoming wavepacket is split at a balanced (50\%-50\%)
input beam splitter (BS1) into two beams that propagate along either
of the two arms of length $L_A$ or $L_B$, then recombine at the 50\%-50\% beam
merger BS2. It follows from Eq.~(\ref{e10c}) that the spin degree of freedom can serve as a peculiar WW detector, since the distance $x$ traversed by the atom is encoded in the spin. Suppose that we know the path lengths $L_A$, $L_{B}$.
We can then guess the path taken by the atom if  we project the atomic internal state on an optimally (with respect to WW measurements) chosen pair of orthogonal spin states, namely
\be
|\tilde \psi'_{A,B} \rangle = 
\frac{1}{\sqrt{2}}|1\rangle \pm 
i \frac{1}{\sqrt{2}} \exp{\left[\rmi\left(\frac{(k_2-k_1)(L_{A}+L_{B})}{2}+\arg\{c_1^*c_2\}\right)\right]}|2\rangle. 
\ee 
In practice, since the exact values of $L_{A}$ and $L_B$ are assumed to be unknown to the 
measuring party, they can use a similar sub-optimal measurement basis 
(defined in terms of $L_{A0}$ and $L_{B0}$, known to them):

\numparts \label{e11}
\be
|\tilde \psi_{A,B}\rangle = 
\frac{1}{\sqrt{2}}|1\rangle \pm 
i \frac{1}{\sqrt{2}} \exp{\left[\rmi\left(\frac{(k_2-k_1)(L_{A0}+L_{B0})}{2}+\arg\{c_1^*c_2\}
\right) \right]}|2\rangle, \label{e11a}
\ee where $c_1,c_2$ are defined in (\ref{e10c}).
 
The internal states of atoms arriving at the beam merger from path $A$ or $B$  
are 
\begin{eqnarray}
\nonumber
&&|\psi_A\rangle = c_1 \exp{(\rmi k_1 L_A)}|1\rangle+
c_2\exp{(\rmi k_2 L_A)}|2\rangle,\\
&&|\psi_B\rangle= c_1 \exp{(\rmi k_1 L_B)}|1\rangle+
c_2 \exp{(\rmi k_2 L_B)}|2\rangle, \label{e10b}
\end{eqnarray}
\endnumparts
respectively. Both detectors ($+$), ($-$) further distinguish between states $|\tilde \psi_{A}\rangle$, $|\tilde \psi_{B}\rangle$ (see Fig.~\ref{f2b}). If we detect (regardless 
by which detector) $|\tilde \psi_{A}\rangle$ ($|\tilde \psi_{B}\rangle$) outside BS2,
we can guess that the atom has followed path $A$ ($B$, respectively). For the antisymmetric case, where $\delta L_B = -\delta L_A\equiv\delta L/2$ (see e.g. \cite{game06}), 
the correct-guess probability is given by 
\numparts
\label{e12}
\be
P_{\rm WW} = |\langle\tilde \psi_{A}|\psi_{A}\rangle |^2 = 
|\langle\tilde \psi_{B}|\psi_{B}\rangle |^2 \equiv
(1+D)/2.
\ee 
It follows from Eqs.~(\ref{e11}) that   
\be
D = 2\sqrt{p_1 p_2}\left| \sin \frac{(k_2-k_1) \left[ \lp L_{AB} \rp_0+ \delta L\right] }{2} \right| =
 2\sqrt{p_1 p_2}\left| \sin \frac{(k_2-k_1) L_{AB} }{2} \right|,
\label{eqD}
\ee
\endnumparts with $p_1=|c_1|^2$, $p_2=|c_2|^2$ and $(L_{AB})_0\equiv L_{A0}-L_{B0}$, $L_{AB}\equiv L_A-L_B$. 
Hence, in contrast to the standard expression (\ref{P_ww}), Eqs. (\ref{e12}) imply that path distinguishability
\emph{oscillates} with $L_{AB}$ for TIE states. This is, in fact, the key reason of all interesting effects described bellow.

We find that $D$ in Eq.~(\ref{eqD}) is the same as that in Eq.~(\ref{e4b}) if we set 
$|d_{A,B}\rangle$ to be $|\psi_{A,B}\rangle$:
\be
D = \sqrt{1-|\langle \psi_B|\psi_A\rangle|^2}.
\label{eqD2}
\ee

\begin{figure}[tbh]
\centering
\centerline{\epsfig{file=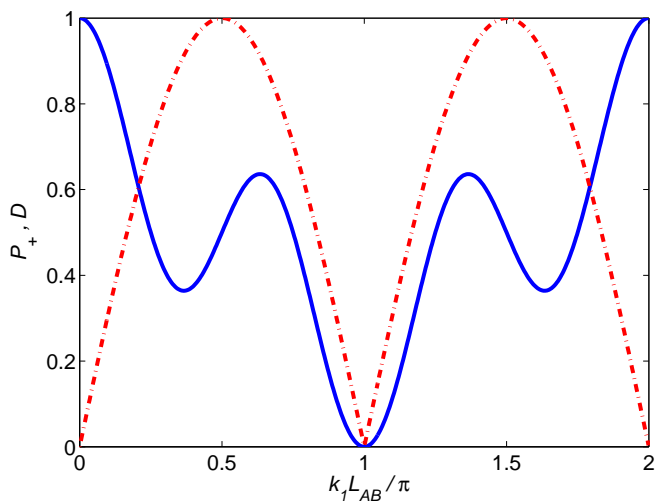,scale=0.5}}

\caption{An example of interference fringes, $P_+$, \eref{intf_pattern} (blue) and path distinguishability \eref{eqD} (red dash-dotted) for TIE states with $k_2=3k_1$.} \label{f2}
\end{figure}

The interference pattern recorded by output detectors ($+$ and $-$) is 
\numparts
\begin{eqnarray} \label{e13}
P_\pm (L_{AB}) &=& p_1P_{1\pm}(L_{AB})+p_2P_{2\pm}(L_{AB}),
\label{intf_pattern} \\
P_{1\pm}(L_{AB})&=&\frac{1}{2}(1 \pm \cos k_1 L_{AB}), \\
P_{2\pm}(L_{AB})&=&\frac{1}{2}(1 \pm \cos k_2 L_{AB}),
\end{eqnarray}
\endnumparts where $P_{1,2\pm}$ are probabilities of hitting a detector by a particle in state $1$ or $2$, respectively. 
As seen from  Fig.~\ref{f2}, this overall non-sinusoidal interference pattern oscillates between $P_+=0$ and $P_+=1$ (and likewise for $P_-$) for TIE states. 
If $k_2 \gg k_1$, then $P_{1\pm}$ and $P_{2\pm}$ are slow-changing and fast-changing, respectively, distinctly from standard states (with $k_1=k_2$).

If $|k_2-k_1| \ll k_{1,2}$, the amount of which-way information in the internal states 
(\ref{e10b}) changes slowly with respect to the interference fringe pattern so that we 
can observe smooth transitions between high-fringe contrast, but low path-distinguishability, 
and those of almost no fringe contrast but high path distinguishability, 
as predicted by (\ref{e1}). Conversely, if $k_1$ and $k_2$ are very different, the 
which-way information varies faster 
than the fringe spacing of the interference pattern which is then, in general, non-sinusoidal.

Let us assume $k_2=3k_1,~~L_{AB}=L_0+\delta L$, where $k_1L_0=\frac{\pi}{2}$ and $|\delta L| \ll |L_0|$, the sign of $\delta L$ being unknown to us. 
Then, analogously to the standard WW and WP probabilities in Sec.~\ref{comp}, we get from 
Eqs.~(\ref{e12}) and (\ref{e13}):
\numparts \label{es14}
\begin{eqnarray}
P_{\rm WW} &=& 1-\frac{\lp k_1\delta L\rp^2}{4} \label{e14a} \\
P_{\rm WP} &=& P_+\lp |\delta L|\rp = P_-\lp -|\delta L|\rp \simeq \frac{1}{2}\lp 1+|k_1\delta L| \rp.  
\end{eqnarray}
\endnumparts
The \emph{linear dependence} of $P_{\rm WP}$ on $|k_1\delta L|$ must be contrasted with the \emph{weaker} dependence of its standard counterpart 
Eq. (\ref{e9}) on $k|\alpha\delta L|$ (see Fig.~\ref{f2}). This shows that inequality (\ref{Duality3}), or equivalently (\ref{e9}),
which holds in cases when $D\neq D(L_{AB})$, cannot be directly applied to the TIE states. Namely, TIE states allow for \emph{better tradeoff} between $P_{\rm WW}$ and $P_{\rm WP}$. 
This remarkable result is operationally unambiguous and definition-free.

\section{Complementarity for TIE}
\label{comptie}
\subsection{Difficulties with visibility definition}

The goal of this subsection is to show that however Ineq.~(\ref{e9}) is strongly violated, relation (\ref{e1}) is not. This means that complementarity in the sense of Eq.~(\ref{e1}) still holds. The reason is that (\ref{e1}) was derived under assumption of $D$ and $P_+$ being dependent on two different (independent) variables. Because our case does not meet this assumption, violation of (\ref{e9}) does not imply violation of (\ref{e1}). We will illustrate now this fact. 

As is clear from Sec.~\ref{model}, scheme plotted on Fig.~\ref{f2} does not allow for {\em independent} changes of distinguishability (\ref{eqD}) and detection probability (\ref{intf_pattern}). Their values are jointly determined by $L_{AB}$, i.e. the arms length difference the particle travels in the magnetic field. To make our setup compatible with assumptions of derivation of Eq.~(\ref{e1}), we can add a magnetic field-free region before the atom enters BS2. In the magnetic field, the spin state of the atom is changed in arm $A$, $B$ differently. Let us denote the final states at the end of this region by $|\psi_A\rangle$, $|\psi_B\rangle$, respectively. After entering field free zone, both spin states have the same potential energy , hence the same wavenumber $k$. The phase difference acquired by the interfering atom right before BS2 is then $\theta=k\overline{L}_{AB}$. Here $\overline{L}_{AB}$ is the length difference of the arms inside the magnetic field-free region, 
and it is independent of $L_{AB}$. In such a modified setup, the overall state right before BS2 is 

\be
|\Psi \rangle = \frac{1}{\sqrt{2}}\left[|\psi_A\rangle |A\rangle + e^{i\theta}|\psi_B\rangle |B\rangle \right].
\ee 

The state of the interfering atom, after tracing out the spin degree of freedom, has the following form in the $|A\rangle,~|B\rangle$ basis:
\numparts
\be
\rho = \frac{1}{2} \left( \begin{array}{cc}
1 & \langle \psi_B|\psi_A \rangle e^{i\theta}\\
\langle \psi_A|\psi_B \rangle e^{-i\theta} & 1
\end{array}\right). \label{eqrho}
\ee

Hence, one can predict the resulting interference pattern 
\be
P_\pm=\frac{1}{2}\left[ 1\pm|\langle \psi_B | \psi_A\rangle|\cos(\theta+\alpha) \right],\;\alpha={\rm Arg}[\langle \psi_B | \psi_A\rangle].\label{st-prob}
\ee
\endnumparts
Calculating the fringe contrast of interference pattern (\ref{st-prob}) yields 
\begin{equation}
\mathcal{V}=|\langle \psi_B | \psi_A\rangle|.
\label{englert_vis}
\end{equation}

Quantity  $|\langle \psi_B | \psi_A\rangle|$ can be identified, as well, with the so called {\em purity} of the state. One can check, that the purity $\mathcal{V}$ (\ref{englert_vis}) and $D$  (\ref{eqD}) fulfill $D^2+\mathcal{V}^2=1$ for all $L_{AB}$. Important fact to realize is that for TIE states, purity, although still complementary to $D$, {\em cannot} be directly interpreted as fringe contrast, i.e. it is not directly connected to the observed interference pattern of the kind (\ref{intf_pattern}). 

Hence, TIE requires different complementary measures characterizing (\ref{intf_pattern}).

\subsection{TIE-state complementarity for discrete WP and WW measurements}

We now turn to a comparison between TIE and ``standard'' schemes as regards
the duality relation (\ref{Duality3})-(\ref{e9}) for \emph{discrete} WP and WW measurements. 
As shown
in Sec.~\ref{WW&WP}, TIE states allow better tradeoff between the two measurements. 
How can it be 
expressed in terms of complementary measures? 

To  bring out the merits of TIE states, propose a quantity directly related to the phase shift measurements via interference fringes, namely, the steepness of the fringe pattern, which we call
the {\em sensitivity,} proportional to the derivative of $P_\pm$, the measured detection probability  with respect to the interferometric phase. 
For standard sinusoidal interference patterns, the intensity, sensitivity and visibility are simply related by
$P_\pm=\frac{1}{2} \pm \frac{V}{2}\cos \phi$, and $S = 2|{\rm d}P_\pm/{\rm d}\phi|=V|\sin \phi|$. Thus, for monochromatic waves it follows from (\ref{e1}) that  the sensitivity
satisfies 
\begin{equation}
S^2+D^2 \le 1. 
\label{e8}
\end{equation}

If we measure small length variations using polychromatic waves,
it is appropriate to relate the phase changes to the \emph{shortest wavelength} used, so that we can define the sensitivity as
\begin{eqnarray}
S= \frac{2}{k_{\rm max}}\left| \frac{{\rm d}P_\pm}{{\rm d}L_{AB}} \right| .
\label{eqS}
\end{eqnarray}

\begin{figure}[tbh]
\centering
\centerline{\epsfig{file=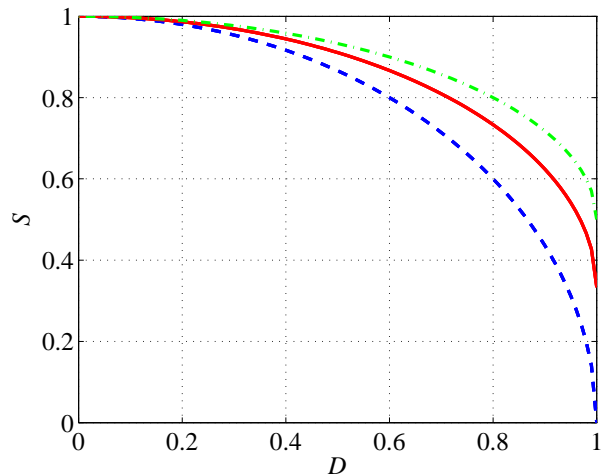,scale=0.6}}
\caption{The upper bounds of $S$-$D$ dependence  for $\kappa=1$ of Eq.~(\ref{e8}) (dashed-blue) and for
$\kappa=3$ (red) and $\kappa\gg 1$ of Eq.~(\ref{eqSN}) (dash-dotted green).} \label{f3}
\end{figure}

If we turn to the state \eref{e2} with $k_2=3k_1$, we find that the sensitivity is $S=\left| \frac{p_1}{3} \sin k_1 L_{AB} + p_2 \sin 3 k_1 L_{AB} \right|$. 
At the points of highest path distinguishability, $L_{AB} = (2n+1)\pi/2k_1$, the sensitivity is  $S=|1-4p_1/3|$.  
The distinguishability at these points is $D=2\sqrt{p_1(1-p_1)}$. By eliminating $p_1$, we find the relation between $S$ and $D$ in the form of an ellipse equation 
(see Fig.~\ref{f3})
\begin{eqnarray}
\left( \frac{S-\frac{1}{3}}{\frac{2}{3}} \right)^2 + D^2 \le 1 .
\label{eqSD}
\end{eqnarray}
This relation, derived for cases when $D=D(L_{AB})$, allows for higher $D$ at a given $S$, and vice versa, than the  complementarity relation \eref{e8} valid in cases $D\neq D(L_{AB})$. This relation can be generalized 
to other ratios between $k_1$ and $k_2 = \kappa k_1$, resulting in
\begin{eqnarray}
\left( \frac{S-\frac{\kappa-1}{2\kappa}}{\frac{\kappa+1}{2\kappa}} \right)^2 + D^2 \le 1 ,
\label{eqSN}
\end{eqnarray}
which in the limit $\kappa\to \infty$ tends to
\begin{eqnarray}
\left( \frac{S-\frac{1}{2}}{\frac{1}{2}} \right)^2 + D^2 \le 1 .
\label{eqSinfty}
\end{eqnarray}
Inequality (\ref{eqSN}) describes the relation between $D$ and maximum $S$ at the point where 
$D(L_{AB})$ is maximum. The equal sign holds for pure WW detector states.
The area enclosed by the ellipse \eref{eqSN} grows with $\kappa$, progressively exceeding the area within the circle \eref{e8}. The \emph{surplus area}
(i.e., the difference between the TIE ellipse area and the standard-complementarity circle area) reflects the \emph{additional} path and phase information provided by the 
TIE state \eref{e2} compared to unentangled states (i.e. for states when all internal states acquire the same interferometric phase).

\section{Which-phase and which-way accuracy bounds} \label{sec6}
\subsection{WW and WP accuracy under shot-noise fluctuations}

We may now compare the bounds on WW and WP accuracy obtainable for TIE cases and cases when $D\neq D(L_{AB})$, so as to determine small changes $\delta L$ of $L_{AB}$. 

We assume that for path lengths $L_{A0}$ and $L_{B0}$
the probability of an atom hitting each detector is 1/2, and
we have maximum distinguishability $D=1$ and sensitivity $S=(\kappa-1)/(2\kappa)$ (i.e., we assume $p_1=1/2$). How many atoms ${\cal N}_{\rm in}$ entering the interferometer
are needed to determine $\delta L$? We must ensure that the change of statistics at each detector, $\delta {\cal N} = {\cal N}_{\rm in} \frac{dP}{dL}\delta L = {\cal N}_{\rm in}
\frac{k_{\rm max}\delta L}{2} S$ is sufficiently bigger than the \emph{shot noise} fluctuation of the number of atoms hitting each detector, 
$\Delta {\cal N} = \frac{\sqrt{{\cal N}_{\rm in}}}{2}$, i.e.,
\begin{eqnarray}
 {\cal N}_{\rm in} > \frac{1}{(k_{\rm max}\delta L S)^2} .
\label{Nin}
\end{eqnarray}
How many among these atoms will have their path inferred incorrectly? This number is closely related to distinguishability, namely ${\cal N}_{\rm wrong}=\frac{1-D}{2}{\cal N}_{\rm in}$. Since the path length difference has been shifted by  $\delta L$, $D$ is no longer equal to unity, but rather $D \approx 1-\frac{\Delta k^2 \delta L^2}{8}$.
This means that the number of atoms with incorrectly inferred path is
\begin{eqnarray}
 {\cal N}_{\rm wrong} > \frac{\Delta k^2}{16 k_{\rm max}^2 S^2} \approx \frac{1}{4},
 \label{Nwrong}
\end{eqnarray}
where we have used the TIE result (\ref{eqSN}) with $\Delta k=\frac{\kappa-1}{\kappa}k_{\rm max}$.

This result can be compared to what we would find in interference experiments without TIE with imperfect path detection. While the necessary number of input atoms is still given by \eref{Nin}, the number of atoms with incorrectly determined path is found from \eref{e8}, yielding 
\begin{eqnarray}
 {\cal N}_{\rm wrong} &=& \frac{1-D}{2}{\cal N}_{\rm in} > 
 \frac{1-\sqrt{1-S^2}}{2} \frac{1}{(k_{\rm max}\delta L S)^2} 
\nonumber \\
&\ge& 
\frac{1}{4k_{\rm max}^2\delta L^2}.\label{Nws} 
\end{eqnarray}
We now witness the tremendous accuracy edge of TIE over the case in which $D\neq D(L_{AB})$: according to \eref{Nin} we would need 90,000 atoms to be sent through the interferometer 
to detect a small length change, e.g., $k_{\rm{max}}\delta L \approx 0.01$ in the TIE scheme with $k_2=3k_1$ and $S=1/3$.
Yet among these atoms, according to \eref{Nwrong}, {\em less than one\/} atom on average will have the path determined incorrectly as opposed
to \emph{thousands} in the standard case \eref{Nws}.
Namely, for the same length change to be detected with high distinguishability $D \geq 0.9$, one would need at
least 53,000 input atoms in the standard
case, out of which we would guess the path incorrectly for about 2,600 atoms. The
number of input atoms would have to increase to infinity if we wish to attain the lowest number of wrong guesses limited by (\ref{Nws}), i.e., 
${\cal N}_{\rm{wrong}}=2,500$.

\subsection{How to verify our guesses?}

A question arises: how to verify the path of the TIE particle we have 
inferred from internal-state measurements? 
This can be done by modifying the setup, i.e., occasionally blocking one of the paths, 
so that if the detector measures internal state $\tilde{\psi}_A$ or $\tilde{\psi}_B$, we know the path of the particle. 
One can conceive of various ``games'' in which one of the parties chooses in certain trials the length shift of the interferometer arms to be $+|\delta L|$ or 
$-|\delta L|$, and in other trials selects the atomic path by blocking the other one. After all trials have been performed, the rival ``guessing party'' is asked to 
determine the length shift and guess which paths the atoms took in the selected subensemble of trials \cite{game06}. The use of TIE states will give the ``guessing party''
a clear edge over someone using non-TIE states in similar games.

\section{Conclusions} \label{sec7}

We have discussed an interferometric scheme employing particles represented by bichromatic waves, whose translational degree of freedom (wave vector) \emph{is entangled} 
with the internal degree of freedom (spin): TIE states. One can then use the internal (spin) state to infer the which-way information. The amount of which-way (WW) 
information varies with the interferometer phase. The usual definition of visibility is not suitable for quantification of wave-particle duality for such entangled states.


Specifically, the cardinal feature of TIE states is that they make the path distinguishability $D$ dependent on the arms-length difference $L_{AB}$ and thus
on the corresponding phase difference, unlike the standard scenario. 
This dependence brings about interesting results, that seemingly violate the standard complementarity relation (\ref{e1}). In Sec.~\ref{comptie} we have shown that this relation is not violated indeed. However, the meaning of some quantities cannot be directly adopted for TIE states. Namely, the purity of the interfering state, $\mathcal{V}$ (\ref{englert_vis}), does not have the direct meaning of contrast of the interference pattern. 

Our ability to combine guessing of an unknown sign of the phase (length) difference $\delta L$ (which-phase--WP guessing) 
with near-certain WW guessing, is \emph{far superior} for TIE compared to the standard case when $D$ is $L_{AB}$ independent.
This may be demonstrated either as the violation of Ineq.~(\ref{Duality3}) constraining $P_{\rm
WW}$ and $P_{\rm WP}$, or as violation of Ineq.~(\ref{e8}) 
constraining the sensitivity $S$ and $D$.
The operational significance of these violations is embodied by the drastically different accuracy limits on the number of wrong path guesses for TIE (Eq.~\ref{Nwrong})
and the standard case (Eq.~\ref{Nws}).
We stress that none of these unusual properties arise for \emph{unentangled} polychromatic waves (Eqs.~\ref{eqs9}), which behave quite standardly.

Although the proposed realization of TIE is challenging, it is within the realm of current possibilities of cold atom manipulation and interferometry. 
We believe that the novel aspects of quantum interferometry associated with TIE warrant this experimental endeavor.

\ack
We would like to thank A. Zeilinger and L. Vaidman for very useful discussions and acknowledge the support of GA\v CR (GA202/05/0486), 
M\v SMT (LC 06007), MSM~(6198959213), GIF and EC (SCALA Network of Excellence).


\end{document}